\documentclass[sigplan,10pt]{acmart}
\AtBeginDocument{%
  }

\usepackage{stmaryrd}
\usepackage{color}
\usepackage{listings}
\usepackage{listingsutf8}
\usepackage{graphicx} 
\usepackage{subcaption} 
\usepackage{booktabs}
\usepackage{tabularx}
\usepackage{array}
\usepackage{lipsum}
\usepackage{tikz}
\usetikzlibrary{arrows.meta}

\definecolor{keywordcolor}{rgb}{0.7, 0.1, 0.1}
\definecolor{tacticcolor}{rgb}{0.0, 0.1, 0.6}
\definecolor{commentcolor}{rgb}{0.4, 0.4, 0.4}
\definecolor{symbolcolor}{rgb}{0.0, 0.1, 0.6}
\definecolor{sortcolor}{rgb}{0.1, 0.5, 0.1}
\definecolor{attributecolor}{rgb}{0.7, 0.1, 0.1}

\setcopyright{acmlicensed}
\copyrightyear{2026}
\acmYear{2026}
\acmDOI{XXXXXXX.XXXXXXX}
\acmConference[Conference acronym 'XX]{Conference title}{June 03--05, 2026}{Woodstock, NY}
\acmISBN{978-1-4503-XXXX-X/2026/09}

\begin{document}

\title{A Lean Paper About Paper: A Formal Framework for Origami}

\author{Celio Boulay, Alexander Chai, Anthony Chang, Thomas Moulin}
\affiliation{%
  \institution{Columbia University in the City of New York}
  \country{New York, NY -- USA}
}

\begin{abstract}
The mathematics of Origami have been well studied and shown to develop several interesting results.
We use Lean 4 tactics and build on Mathlib to redefine the 7 Huzita operations as theorems instead of axioms and prove their existence. We develop proofs for important origami constructions (such as trisecting an angle), implement origami-constructible numbers and prove the associated Cardano's formula, and formalize Haga's theorem. A \textit{Crease Pattern Inspector} explores physical folding by providing a full pipeline to create and visualize models constrained by the Huzita formalism. The Lean codebase brings 100+ theorems and lemmas.

\end{abstract}



\maketitle

\section{Introduction}

While most work in the Lean programming language focuses on theoretical results, formal proofs have made their way into the \textit{real world} (e.g., AWS Clean Rooms with SampCert for differential privacy \cite{samp_cert_aws}). Embedding Lean into the physical world improves soundness, an essential aspect of autonomous robots and self-driving vehicles \cite{kent_galois} among others. The leap from theory to the physical world is not straightforward, which is why we chose to study the mathematics of origami, giving an excellent framework for this purpose, and grounded in the real world (most of the theory can be replicated by folding a piece of paper). Physical origami can be analyzed from the crease pattern from which many properties can be derived. For algebraic considerations, viewing origami as sequences of constrained axiomatic operations enables mathematical constructions beyond what is possible using compass-and-straightedge.

\subsection{Related Work}
\label{sec:related-work}

Mathematical origami has been studied both as a theory of geometric
constructions and as a theory of folded-sheet configurations. On the
construction side, the standard single-fold model is usually presented through
the Huzita--Justin (also known as Huzita--Hatori or simply Huzita) operations. Huzita's formulation of
the first six operations \cite{Huzita1989,Huzita1992} and Justin's independent
enumeration of the seven point--line alignment cases \cite{Justin1989} provide
the historical basis for this terminology; the seventh operation was later
rediscovered by Hatori \cite{AlperinLang2009}. The status of these rules is subtle: they are best
viewed as a catalogue of elementary alignment operations subject to appropriate
non-degeneracy and solvability hypotheses, rather than as unconditional
existence-and-uniqueness assertions. In particular, some operations may admit
several folds or none in degenerate configurations. Alperin and Lang give a
systematic account of the completeness of the single-fold alignment cases and
of extensions to multi-fold operations \cite{AlperinLang2009}; Kasem, Ghourabi,
and Ida likewise emphasize the need to state the exceptional cases precisely
when treating the operations as logical axioms \cite{KasemGhourabiIda2011}.

The construction model is algebraically more expressive than straightedge and
compass. The simultaneous point-to-line operation (usually called O6 or the
Beloch fold) is associated with a cubic constraint, explaining classical
origami constructions such as angle trisection. This connection goes back to
Beloch's work on solving cubics by folding and is developed in modern
field-theoretic terms by Alperin \cite{Beloch1936,Alperin2000}. In particular,
origami-constructible numbers form a field that can equivalently be described
through intersections of conics or a marked ruler \cite{Alperin2000}. More
recent logical work develops first-order axiom systems in the spirit of
Huzita--Justin and relates their constructible-point models to suitable field
theories \cite{BeklemishevDmitrievaMakowsky2024}. These results motivate
retaining the geometric content of the individual fold operations explicitly,
rather than reducing a construction immediately to a numerical solver.

At the level of concrete paper subdivisions, Haga's theorems provide
particularly useful exact constructions. Starting from a square or rectangle,
suitable corner-to-edge folds generate division points such as thirds without
measurement; more generally, Haga-style constructions supply rational reference
points that are ubiquitous in crease-pattern design \cite{Haga2008}. This
perspective is directly reflected in our development: a formalization of
Haga's first theorem derives the relevant incidence statement from the
reflection induced by a fold. It therefore forms a useful bridge between an
elementary paper-folding construction and the coordinate-level representation
used by the system.

Computational-origami systems have pursued this explicit route in a largely
symbolic setting. Ghourabi, Ida, Takahashi, Marin, and Kasem express Huzita's
operations as first-order geometric constraints, translate them into polynomial
equalities, disequalities, and inequalities, and use this translation for
construction, visualization, and automated reasoning \cite{GhourabiEtAl2007}.
The Eos line of work subsequently combined such construction models with
algebraic proving; for example, Ida, Ghourabi, and Takahashi give
computer-assisted constructions and Gröbner-basis-based verification for
polygonal knot origami \cite{IdaGhourabiTakahashi2015}. In a closer precursor
to proof-assistant work, Kaliszyk and Ida studied how decision procedures,
notably algebraic ones, can support machine-checked proofs about origami
constructions in Coq and Isabelle/HOL \cite{KaliszykIda2011}. Their analysis
also highlights the difficulty of bridging geometric fold predicates,
algebraic constraints, and the branching solution sets of the non-linear
operations.

Our development takes a complementary Lean~4-oriented approach. At the
construction level, points and normalized lines are represented by real
coordinates, folds act by reflection, and the seven Huzita operations
are exposed as typed interfaces. This supports replaying a sequence of
user-selected folds as a Lean term and proving concrete geometric consequences;
the coordinate-level trisection identity used in the treatment of Abe's
acute-angle construction is an example of the latter. At the configuration level, the development uses rational-coordinate
vertices and triangular faces, together with an explicit face map, overlap
predicate, and superposition relation. A fold step is required to preserve the
face correspondence and update layer order coherently. This separation makes explicit a distinction often hidden in construction-focused accounts: producing a crease satisfying an alignment constraint is not yet a proof that a finite sheet can execute the fold without inconsistent layering or self-intersection.

That distinction is central in computational geometry. Layer order and
self-intersection are substantive global constraints even in flat-folding,
while related rigid-foldability problems are computationally intractable in
general \cite{AkitayaEtAl2020}. Consequently, the present formalization is not
intended to replace physical or rigid-origami simulation. Instead, it supplies
a proof-assistant substrate on which the two levels, axiomatic construction
and state-transition validity, can be connected incrementally. An important
next step is to discharge the fold-operation interfaces from the coordinate
definitions under precise side conditions, and to connect the generated
construction traces to verified configuration transitions.
\subsection{Paper Outline}

The subsequent text of this paper is organized as follows:
Section \ref{sec:implementation} outlines our methodology and process with subsection \ref{ssec:structure} covering the structures we created and choices we made in representing origami folds and models, and subsection \ref{ssec:axioms} covering the Huzita Axioms.
Section \ref{sec:objects} outlines our key results, including proofs of preexisting results that we formalized in Lean and objects and tools we created.
Section \ref{sec:conclusion} discusses the conclusions of our work as well as avenues for further inquiry.
Finally, we conclude this paper with our AI, Ethics, and Privacy disclosures, acknowledgments, as well as our references.
\section{Implementation}
\label{sec:implementation}
When describing origami constructions formally, a natural framework is given by the Huzita axioms.
Rather than treating an origami construction as a sequence of physical folds, these axioms characterize the geometric operations that can be performed on a sheet of paper.
Therefore they provide a concise and precise language for reasoning about origami.
Our implementation therefore takes the Huzita axioms as its basic interface for origami.
Throughout this section we will talk about our implementation and the choices we made.
\subsection{Structure}
\label{ssec:structure}
Before discussing the Huzita Folds~\ref{ssec:axioms}, we need to fix our representation of the underlying geometry.
We fix a line by the coefficients of the equation $ax + by + c = 0$.
However since we not only prove the existence but also the uniqueness of some of the Huzita folds we must deal with the natural issue of different coefficients representing the same line.
A few natural solutions arise.
We can declare all proportional triples to be the same object so $(1,2,3)$ and $(2,4,6)$ are equal by definition.
But then no statements/properties afterwards may distinguish proportional triples.
We can retain all triples and, rather than asserting that two lines are equal, state that their coefficients are proportional.
The resulting scale factor then appears in every statement and must be carried consistently throughout the proof.
Or we can allow only one triple per line so plain equality already means geometric equality.
We opted for the last option.
The other two options allow equality of lines cheaply but burden every function we later define with a proof that it is well defined with respect to the choice of the representative.
Normalization pays this cost once at construction time.
\begin{lstlisting}
structure Line where
  a : ℝ
  b : ℝ
  c : ℝ
  nontrivial : a ≠ 0 ∨ b ≠ 0
  normalized : a = 1 ∨ a = 0 ∧ b = 1
\end{lstlisting}
\subsection{Huzita Folds}\label{ssec:axioms}
With lines represented explicitly, each folding instruction can now be translated into a precise geometric constraint on those lines.
We identify a fold with its crease and interpret placement as reflection across that line.
Thus, while they are referred to as axioms, the Huzita folds are proven as a theorem of our coordinate model rather than as additional assumptions.
Note that these axioms are not physically bounded.
\tikzset{
  hz line/.style={line width=0.65pt},
  hz crease/.style={line width=0.75pt,dash pattern=on 3pt off 1.8pt},
  hz aux/.style={line width=0.45pt,draw=black!48},
  hz arrow/.style={line width=0.5pt,shorten >=2pt,-{Stealth[length=3.5pt,width=3pt]}},
  hz label/.style={inner sep=1.5pt},
  hz point/.style={circle,fill=black,inner sep=1.15pt},
  hz image/.style={circle,draw=black,fill=white,line width=0.55pt,inner sep=1.3pt}
}
\begin{figure*}[t]
\centering
\begin{tikzpicture}[x=1cm,y=1cm,font=\small,line cap=round,line join=round]
\path[use as bounding box] (-2.1,-4.35) rectangle (14.7,1.45);
\begin{scope}
  \node[font=\small\bfseries] at (0,1.23) {H1. Through two points};
  \draw[hz crease] (-1.45,-0.64444)--(1.45,0.64444);
  \node[hz label,right] at (1.43,0.64) {$f$};
  \node[hz point] at (-0.9,-0.4) {};
  \node[hz label,above left] at (-0.9,-0.4) {$p_1$};
  \node[hz point] at (0.9,0.4) {};
  \node[hz label,above left] at (0.9,0.4) {$p_2$};
  \node at (0,-1.17) {$p_1,p_2\in f$};
\end{scope}
\begin{scope}[xshift=4.2cm]
  \node[font=\small\bfseries] at (0,1.23) {H2. Point onto point};
  \draw[hz crease] (0,-0.84)--(0,0.84);
  \node[hz label,right] at (0,0.81) {$f$};
  \draw[hz arrow] (-1.05,0)--(1.05,0);
  \draw[hz aux] (0,0.15)--(0.15,0.15)--(0.15,0);
  \node[hz point] at (-1.05,0) {};
  \node[hz point] at (1.05,0) {};
  \node[hz label,above] at (-1.05,0) {$p_1$};
  \node[hz label,above] at (1.05,0) {$p_2$};
  \node at (0,-1.17) {$r_f(p_1)=p_2$};
\end{scope}
\begin{scope}[xshift=8.4cm]
  \node[font=\small\bfseries] at (0,1.23) {H3. Line onto line};
  \draw[hz line] (-1.2,0)--(1.2,0) node[hz label,right] {$l_1$};
  \draw[hz line] (0,-0.9)--(0,0.9) node[hz label,right] {$l_2$};
  \draw[hz crease] (-0.84,-0.84)--(0.84,0.84);
  \draw[hz crease] (-0.84,0.84)--(0.84,-0.84);
  \node[hz label,right] at (0.85,0.8) {$f$};
  \node[hz label,right] at (0.85,-0.8) {$g$};
  \node at (0,-1.17) {$r_f(l_1)=r_g(l_1)=l_2$};
\end{scope}
\begin{scope}[xshift=12.6cm]
  \node[font=\small\bfseries] at (0,1.23) {H4. Perpendicular through a point};
  \draw[hz line] (-1.3,-0.32)--(1.25,-0.32) node[hz label,right] {$l$};
  \draw[hz crease] (0.2,-0.85)--(0.2,0.91);
  \node[hz label,right] at (0.2,-0.78) {$f$};
  \draw[hz aux] (0.2,-0.15)--(0.37,-0.15)--(0.37,-0.32);
  \node[hz point] at (0.2,0.53) {};
  \node[hz label,right] at (0.2,0.53) {$p$};
  \node at (0,-1.17) {$p\in f,\quad f\perp l$};
\end{scope}
\begin{scope}[yshift=-2.95cm]
  \node[font=\small\bfseries] at (0,1.23) {H5. Incidence and placement};
  \draw[hz line] (-0.25,0.875)--(1.5,0);
  \node[hz label,below right] at (1.4,0.05) {$l_1$};
  \draw[hz crease] (0,-0.85)--(0,0.91);
  \node[hz label,right] at (0,-0.84) {$f$};
  \draw[hz arrow] (-1,0.25)--(1,0.25);
  \node[hz point] at (-1,0.25) {};
  \node[hz label,above] at (-1,0.25) {$p_1$};
  \node[hz image] at (1,0.25) {};
  \node[hz label,above right] at (1,0.25) {$q$};
  \node[hz point] at (0,-0.58) {};
  \node[hz label,left] at (0,-0.58) {$p_2$};
  \node at (0,-1.17) {$p_2\in f,\quad r_f(p_1)\in l_1$};
\end{scope}
\begin{scope}[xshift=4.2cm,yshift=-2.95cm]
  \node[font=\small\bfseries] at (0,1.23) {H6. Two point placements};
  \draw[hz line] (1,0.03)--(1,0.94);
  \node[hz label,right] at (1,0.87) {$l_1$};
  \draw[hz line] (0.25,-0.734)--(1.44,-0.1985) node[hz label,right] {$l_2$};
  \draw[hz crease] (0,-0.9)--(0,0.96);
  \node[hz label,left] at (0,0.85) {$f$};
  \draw[hz arrow] (-1,0.5)--(1,0.5);
  \draw[hz arrow] (-0.77,-0.5)--(0.77,-0.5);
  \node[hz point] at (-1,0.5) {};
  \node[hz label,above] at (-1,0.5) {$p_1$};
  \node[hz point] at (-0.77,-0.5) {};
  \node[hz label,below] at (-0.77,-0.5) {$p_2$};
  \node[hz image] at (1,0.5) {};
  \node[hz label,right,xshift=1.5pt] at (1,0.5) {$q_1$};
  \node[hz image] at (0.77,-0.5) {};
  \node[hz label,below] at (0.77,-0.5) {$q_2$};
  \node at (0,-1.17) {$r_f(p_i)\in l_i\quad(i=1,2)$};
\end{scope}
\begin{scope}[xshift=8.4cm,yshift=-2.95cm]
  \node[font=\small\bfseries] at (0,1.23) {H7. Perpendicular placement};
  \draw[hz line] (1,-0.05)--(1,0.94);
  \node[hz label,right] at (1,0.85) {$l_1$};
  \draw[hz line] (-1.22,-0.57)--(1.25,-0.57) node[hz label,right] {$l_2$};
  \draw[hz crease] (0,-0.9)--(0,0.96);
  \node[hz label,left] at (0,0.85) {$f$};
  \draw[hz aux] (0,-0.4)--(0.17,-0.4)--(0.17,-0.57);
  \draw[hz arrow] (-1,0.28)--(1,0.28);
  \node[hz point] at (-1,0.28) {};
  \node[hz label,above] at (-1,0.28) {$p$};
  \node[hz image] at (1,0.28) {};
  \node[hz label,right,xshift=1.5pt] at (1,0.28) {$q$};
  \node at (0,-1.17) {$r_f(p)\in l_1,\quad f\perp l_2$};
\end{scope}
\begin{scope}[xshift=12.6cm,yshift=-2.95cm]
  \node[font=\small\bfseries] at (0,1.23) {Reading the diagrams};
  \draw[hz line] (-1.42,0.75)--(-0.65,0.75);
  \node[anchor=west] at (-0.48,0.75) {given line};
  \draw[hz crease] (-1.42,0.29)--(-0.65,0.29);
  \node[anchor=west] at (-0.48,0.29) {crease};
  \node[hz point] at (-1.04,-0.17) {};
  \node[anchor=west] at (-0.48,-0.17) {given point};
  \node[hz image] at (-1.04,-0.63) {};
  \node[anchor=west] at (-0.48,-0.63) {image point};
  \draw[hz arrow] (-1.42,-1.09)--(-0.65,-1.09);
  \node[anchor=west] at (-0.48,-1.09) {reflection};
\end{scope}
\end{tikzpicture}
\caption{The seven geometric constraints. Each panel shows a satisfying crease, except H3, which shows two distinct solutions for the same inputs.} 
\label{fig:huzita-overview}
\end{figure*}
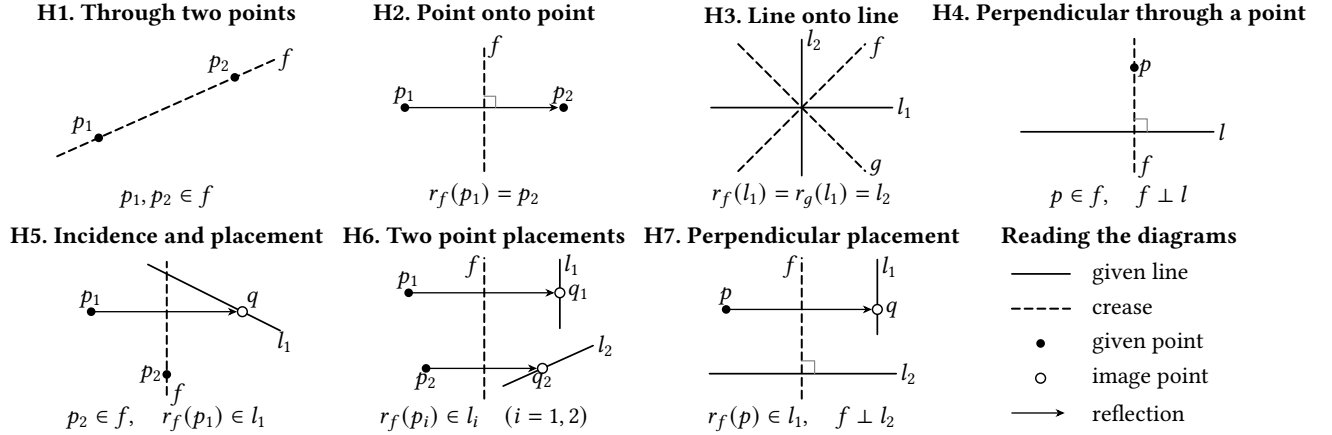
\par
Figure~\ref{fig:huzita-overview} shows the constraints of the seven operations.
The distinction between lying on a crease and being placed onto a target is important, \texttt{f\_places\_p} returns the reflected point.
Writing $r_f$ for reflection, placement of $p$ onto $l$ means $r_f(p) \in l$. The latter incidence condition is written \texttt{on\_line} in Lean.
The operation \texttt{f\_places\_l} returns the reflected line denoted as $r_f(l)$.
For a crease $ax + by + c= 0$, the point reflection is
\[r_f(x,y) = (x,y) - 2 \frac{ax+by + c}{a^2 + b^2}(a,b).\]
To ensure that the denominator is nonzero we similarly require it to have a nonzero-normal.
We prove that reflection fixes exactly the points on the crease, preserves squared distances, and is its own inverse.
Point and line reflection are connected by
\[p \in l \Longleftrightarrow r_f(p) \in r_f(l).\]
Together with the inverse property, this shows that the computed reflected line contains exactly the reflected points of the original line.
\par
The instructions do not all determine a unique crease, or even admit a crease for arbitrary inputs.
Table~\ref{tab:huzita-scope} records the hypotheses and conclusions we prove.
Parallelism includes coincident lines since our definition compares their normal vectors.
\begin{table}[htbp]
  \caption{Here $d$ denotes Euclidean distance; $\exists!$ denotes unique existence.}
  \label{tab:huzita-scope}

  \centering
  \small
  \setlength{\tabcolsep}{5pt}
  \renewcommand{\arraystretch}{1.12}

  \begin{tabularx}{\columnwidth}{
    @{}l >{\raggedright\arraybackslash}X c@{}
  }
    \toprule
    Fold & Hypothesis & Conclusion \\
    \midrule
    H1 & $p_1 \ne p_2$ & $\exists!$ \\
    H2 & $p_1 \ne p_2$ & $\exists!$ \\
    H3 & None & $\exists$ \\
    H4 & None & $\exists!$ \\
    H5 & $d(p_2,l_1)^2 \le d(p_1,p_2)^2$ & $\exists$ \\
    H6 & $l_1 \not\parallel l_2$ & $\exists$ \\
    H7 & $l_1 \not\parallel l_2$ & $\exists!$ \\
    \bottomrule
  \end{tabularx}
\end{table}
\paragraph{Construction, correctness, and equality.}
For Huzita Axioms 1-4 and 7, we define crease-valued functions and prove their properties separately.
For the rest we produce witnesses inside existence proofs.
Since many folds will rely on the property of earlier constructions this will prevent us from needing to repeat their proofs.
Here "construction" means a witness over $\mathbb{R}$, not an executable procedure.
\par
The calculations use convenient coefficients before normalization.
The constructor \texttt{mk\_line} converts these into a normalized line once the normal is shown to be nonzero.
Supporting lemmas allow incidence and reflection to be checked using the original coefficients so the proofs don't have to repeatedly unfold the normalization.
For uniqueness we show that any other solution has proportional coefficients.
\paragraph{Huzita 1:} Given two distinct points $p_1$ and $p_2$ there is a unique fold that passes through both of them.
\begin{lstlisting}
theorem huzita_1 (p₁ p₂ : Point) (h : p₁ ≠ p₂) :
  ∃! f : Fold, f_through_p f p₁ ∧ f_through_p f p₂
\end{lstlisting}
Let $p_i = (x_i, y_i)$, the constructed crease uses
\[A = y_2 - y_1, \qquad B = x_1 - x_2 \qquad C = x_2y_1-x_1y_2.\]
Unlike a slope-based formula, this does not divide by $x_2 - x_1$, so vertical lines need no separate construction.
Distinctness ensures that $A$ and $B$ are not both zero.
Substitution of either point into $Ax + By + C = 0$ gives a polynomial identity, and the corresponding formal checks are closed by \texttt{ring}.
For uniqueness, subtracting the two incidence equations of another solution shows that its first two coefficients are proportional to $(A,B)$.
Substituting back gives the same factor for the constant coefficient.
Normalization therefore identifies the two creases.
\paragraph{Huzita 2:} Given two distinct points there is a fold that places $p_1$ onto $p_2$.
\begin{lstlisting}
theorem huzita_2 (p₁ p₂ : Point) (h : p₁ ≠ p₂) :
  ∃! f : Fold, f_places_p f p₁ = p₂
\end{lstlisting}
We construct the perpendicular bisector by taking the displacement between the points as its normal and choosing the constant coefficient so that it passes through their midpoint
\[(A,B) = (x_2 - x_1, y_2 - y_1), \qquad C = \frac{x_1^2 + y_1^2 - x_2^2-y^2_2}{2}.\]
The distinctness condition makes $A^2 + B^2$ nonzero.
Substitution into the reflection formula, followed by clearing this denominator will verify placement.
Conversely, the reflection equations force any solution to have this direction and contain the midpoint, determining its coefficients up to scale.
Coincident points would again destroy uniqueness, since any crease through the common point fixes it.
\par
The construction also comes with an equal distance characterization.
Writing $f$ for the constructed perpendicular bisector, we prove
\[q \in f \Longleftrightarrow d(q,p_1)^2 = d(q,p_2)^2. \]
Huzita 5 will later utilize this characterization after choosing an image point on its target line.
It shows that the perpendicular bisector of the original point and its image also contains the required fixed point, without repeating the coordinate proof.
\paragraph{Huzita 3:} Given two lines $l_1$ and $l_2$, there is a fold that places $l_1$ onto $l_2$.
\begin{lstlisting}
theorem huzita_3 (l₁ l₂ : Line) :
  ∃ f : Fold, f_places_l f l₁ = l₂
\end{lstlisting}
The two diagonal creases in Figure~\ref{fig:huzita-overview} both reflect the horizontal axis onto the vertical axis.
We verify this example formally, so the theorem deliberately asserts existence rather than unique existence.
\par
\begin{figure}[t]
  \centering

  \tikzset{
    hz line/.style={
      line width=0.65pt
    },
    hz crease/.style={
      line width=0.75pt,
      dash pattern=on 3pt off 1.8pt
    },
    hz aux/.style={
      line width=0.45pt,
      draw=black!48
    },
    hz arrow/.style={
      line width=0.5pt,
      shorten >=2pt,
      -{Stealth[length=3.5pt,width=3pt]}
    },
    hz label/.style={
      inner sep=1.5pt
    },
    hz point/.style={
      circle,
      fill=black,
      inner sep=1.15pt
    },
    hz image/.style={
      circle,
      draw=black,
      fill=white,
      line width=0.55pt,
      inner sep=1.3pt
    }
  }

  \providecommand{\Description}[1]{}

  \resizebox{\linewidth}{!}{%
    \begin{tikzpicture}[
      x=1cm,
      y=1cm,
      font=\small,
      line cap=round,
      line join=round
    ]
      \path[use as bounding box]
        (0,0) rectangle (8.25,4.0);

      \node[font=\small\bfseries] at (2.0,3.83)
        {(a) Construct the crease};
      \node[font=\small\bfseries] at (6.35,3.83)
        {(b) Can an image be chosen?};

      \begin{scope}[
        shift={(1.95,1.73)},
        x=0.88cm,
        y=0.88cm
      ]
        \coordinate (P) at (-1.2,0.9);
        \coordinate (O) at (0,0);
        \coordinate (Q) at (0.9,1.2);

        \draw[hz aux] (O) circle (1.5);
        \draw[hz line] (-1.85,1.2) -- (1.84,1.2);
        \node[hz label,right] at (1.84,1.2) {$l_1$};

        \draw[hz crease] (0.24,-1.68) -- (-0.255,1.785);
        \node[hz label,right] at (0.23,-1.61) {$f$};

        \draw[hz aux] (P) -- (O) -- (Q);
        \node[fill=white,inner sep=1pt] at (0.58,0.51) {$R$};
        \draw[hz arrow] (P) -- (Q);

        \draw[hz aux]
          (-0.00151,1.07121)
          -- (0.01970,0.92272)
          -- (-0.12879,0.90151);

        \node[hz point] at (P) {};
        \node[hz label,above left] at (P) {$p_1$};

        \node[hz image] at (Q) {};
        \node[hz label,above right] at (Q) {$q$};

        \node[hz point] at (O) {};
        \node[hz label,below left] at (O) {$p_2$};

        \node[hz image] at (-0.9,1.2) {};
      \end{scope}

      \foreach \yy/\hh/\cond/\outcome in {
        2.93/0.17/{\delta<R}/{two images},
        1.84/0.4/{\delta=R}/{one image},
        0.75/0.61/{\delta>R}/{no image}
      } {
        \draw[hz aux] (5.28,\yy) circle (0.4);
        \node[hz point,inner sep=0.65pt] at (5.28,\yy) {};

        \draw[hz line]
          (4.65,\yy+\hh) -- (5.91,\yy+\hh);

        \node[anchor=west] at (6.1,\yy+0.1) {$\cond$};
        \node[anchor=west,font=\footnotesize]
          at (6.1,\yy-0.22) {\outcome};
      }

      \node[hz image,inner sep=0.95pt] at (4.91793,3.10) {};
      \node[hz image,inner sep=0.95pt] at (5.64207,3.10) {};
      \node[hz image,inner sep=0.95pt] at (5.28,2.24) {};
    \end{tikzpicture}%
  }

  \caption{Huzita 5's axiom}

  \label{fig:huzita-five}
\end{figure}
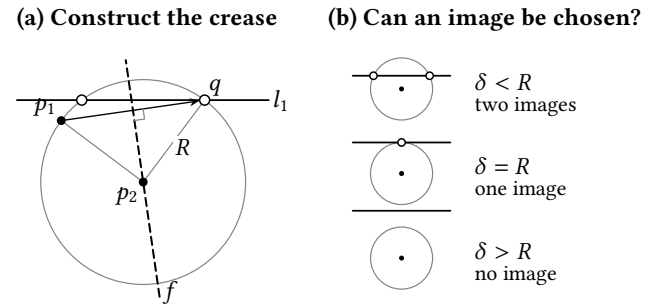

For input equations $a_i x + b_i y + c_i = 0$, let $s_i = \sqrt{a_i^2 + b_i^2}$. We construct one crease from
\[(A,B,C) = s_2(a_1, b_1, c_1) + s_1(a_2, b_2, c_2).\]
This selects an angle bisector when the lines intersect.
For distinct parallel lines it gives the line midway between them; for coincident lines it returns the common line itself.
These are descriptions of the chosen witness, not additional claims that every input has a unique solution.
Here normalization has a second benefit: a stored normal has first coordinate 1, or is (0,1).
The two normals cannot point in opposite directions, so their positive weighted sum cannot vanish. The reflection calculation is proved first for weights whose squares equal the squared normal lengths, assuming that the resulting normal is nonzero, and then applied to the positive square roots.
\paragraph{Huzita 4:} Given a point $p_1$ and a line $l_1$, there is a unique fold perpendicular to $l_1$ that passes through $p_1$.
\begin{lstlisting}
theorem huzita_4 (p : Point) (l : Line) :
  ∃! f : Fold,
    perpendicular f l ∧ f_through_p f p
\end{lstlisting}
For a line with normal $(a,b)$ and a point $p = (x_0, y_0)$, we use $(-b, a, bx_0 - ay_0)$ as the crease coefficients.
The input normal is already nonzero, so this always defines a line.
Perpendicularity determines the direction and incidence determines the position, giving uniqueness.
Huzita 5 and Huzita 6 can reuse this construction in fixed-point cases. When the relevant points coincide and the common point already lies on each required target, a crease through that point suffices.
Coincidence alone does not justify fixing an off-target point.
\paragraph{Huzita 5:}
Given two points $p_1$ and $p_2$ and a line $l_1$, there is a fold that places $p_1$ onto $l_1$ and passes through $p_2$.
Figure~\ref{fig:huzita-five} explains the construction and existence condition better.
Because reflection fixes $p_2$ and preserves distance, the image $q = r_f(p_1)$ must lie on the target line and on the circle centered at $p_2$ with radius $d(p_1, p_2)$.
Conversely, such an image point can be used to construct a fold.
Our theorem statement is written as,
\begin{lstlisting}
theorem huzita_5 (p₁ p₂ : Point) (l₁ : Line)
  (h : dist2_line l₁ p₂ ≤ dist2 p₁ p₂) :
  ∃ f : Fold,
    f_through_p f p₂ ∧
    on_line l₁ (f_places_p f p₁)
\end{lstlisting}
Here \texttt{dist2} measures squared point to point distance.
For a line $l$ with coefficients $(a_l, b_l, c_l)$, \texttt{dist2\_line} gives the squared point to line distance
\[d((x,y), l)^2 = \frac{(a_l x+b_l y + c_l)^2}{a_l^2 + b_l^2}.\]
Using squared distances avoids square roots in the hypothesis and distance comparisons. 
The intersection construction still needs a square root.
The non strict inequality guarantees a nonnegative radicand and includes tangency.
\par
We state the intersection result directly as $q \in l_1$ and $d(p_2, q)^2 = d(p_1, p_2)^2$, rather than introducing a circle into the formal statement.
The helper constructs $q$ by projecting $p_2$ onto the target line and adding a suitable offset along that line.
Only one intersection point is needed, rather than an enumeration of all intersections.
When $q \neq p_1$, Huzita 2 supplies the perpendicular bisector of $p_1 q$.
Its placement property sends $p_1$ to $q$, and its equal-distance characterization shows that it contains $p_2$.
Thus the two requirements follow from properties of an earlier construction.
\par
The case $q = p_1$ must be treated separately, the perpendicular bisector construction requires distinct points.
Here $p_1$ already lies on its target.
For distinct $p_1$, $p_2$, Huzita 1 supplies a crease through both and hence fixes $p_1$.
If the points also coincide, Huzita 4 supplies a crease through their common location.
Thus the theorem includes the zero-radius case without a distinctness assumption, if $p_1 = p_2 = p$, it is solvable exactly when $p \in l_1$, and then every crease through $p$ works.
There is only one image, $p$ but there are multiple creases.
This is the exception to the nondegenerate picture in Fig \ref{fig:huzita-five}.
For necessity, any admissible image $q$ lies on $l_1$, so $d(p_2, l_1) \leq d(p_2, q) = d(p_1, p_2)$ by distance preservation.
Thus the hypothesis characterizes existence exactly.
The converse is packaged with Huzita 5 in \texttt{huzita\_5\_iff}
\[(\exists f, p_2 \in f \vee r_f(p_1) \in l_1) \Longleftrightarrow d(p_2, l_1)^2 \leq d(p_1, p_2)^2.\]
\paragraph{Huzita 6:}
Given two points $p_1$ and $p_2$ and two lines $l_1$ and $l_2$, there is a fold that places $p_1$, onto $l_1$, and $p_2$ onto $l_2$.
\begin{lstlisting}
theorem huzita_6 (p₁ p₂ : Point) (l₁ l₂ : Line)
  (h : ¬ parallel l₁ l₂) :
  ∃ f : Fold,
    on_line l₁ (f_places_p f p₁) ∧
    on_line l₂ (f_places_p f p₂)
\end{lstlisting}
Here the main step is not a closed coordinate formula but an existence argument for a polynomial root.
We first find a nonzero normal $(a,b)$ and then recover the constant coefficient $c$.
This separates finding a crease direction from checking that it gives a solution to the original placement problem.
\par
For $p_i=(x_i,y_i)$ and target equations $\alpha_i x+\beta_i y+\gamma_i=0$, write
\[
  D_i=\alpha_i x_i+\beta_i y_i+\gamma_i,
  \qquad L_i=\alpha_i a+\beta_i b.
\]
Thus $D_i$ is the target equation evaluated at $p_i$, and $L_i$ is the dot product of the target normal and the proposed crease normal.
The reflection formula gives
\[
  (a^2+b^2)D_i=2L_i(ax_i+by_i+c),\qquad i=1,2.
\]
Eliminating $c$ yields
\begin{equation}
    (a^2+b^2)(L_1D_2-L_2D_1)=2L_1L_2\bigl(a(x_2-x_1)+b(y_2-y_1)\bigr).
  \label{eq:huzita-cubic}
\end{equation}
The expression is homogeneous of degree three in $a,b$, although it may vanish identically for particular inputs.
We therefore seek a nonzero solution of
\[
  u a^3+v a^2b+w ab^2+z b^3=0,
  \qquad (a,b)\ne(0,0).
\]
Keeping both coordinates avoids excluding directions with $b=0$.
If $u=0$, $(1,0)$ is already a solution.
Otherwise we set $b=1$ and obtain an ordinary cubic with nonzero leading coefficient.
We prove that it has a real root using the intermediate value theorem, with explicit bounds at which its values have opposite signs.
For $u>0$, the proof uses $M=(|v|+|w|+|z|)/u+1$ and the interval $[-M,M]$; a negative leading coefficient is handled by negating the polynomial.
This establishes existence without providing a root-finding procedure or enumerating the roots.
\par
A solution of the cubic must still give a valid crease.
First suppose $p_1\notin l_1$, so $D_1\ne0$.
If $L_1=0$, Equation~\eqref{eq:huzita-cubic} forces $L_2=0$, since $a^2+b^2\ne0$ and $D_1\ne0$.
A nonzero crease normal would then be perpendicular to both target normals, contradicting nonparallelism.
Consequently we can set
\[
  c=\frac{(a^2+b^2)D_1}{2L_1}-ax_1-by_1.
\]
The first placement equation holds by construction and the cubic implies the second.
This step is essential: finding a root after elimination would not, by itself, justify division by $L_1$ or guarantee the original constraints.
\par
If only the second point lies off its target, we exchange the two point line pairs.
If both are already on their targets, a crease fixing both is sufficient; Huzita 1 handles distinct points and Huzita 4 handles coincident points.
Nonparallelism is sufficient, but not necessary for every solvable instance.
For example, the development verifies that $p_1=(0,1)$ and $p_2=(0,2)$ cannot be placed onto $l_1:\ y=0$ and $l_2:\ y=-2$, respectively.
Geometrically, their images would be at least two units apart, contradicting preservation of their one-unit separation; the formal counterexample is checked algebraically.
Thus, the hypothesis cannot simply be removed from the universal existence statement.
We do not claim uniqueness or classify all H6 solutions, including which parallel configurations are solvable.
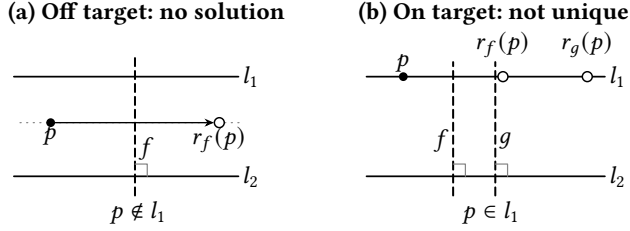
\begin{figure}[t]
  \centering

  \tikzset{
    hz line/.style={
      line width=0.65pt
    },
    hz crease/.style={
      line width=0.75pt,
      dash pattern=on 3pt off 1.8pt
    },
    hz aux/.style={
      line width=0.45pt,
      draw=black!48
    },
    hz arrow/.style={
      line width=0.5pt,
      shorten >=2pt,
      -{Stealth[length=3.5pt,width=3pt]}
    },
    hz label/.style={
      inner sep=1.5pt
    },
    hz point/.style={
      circle,
      fill=black,
      inner sep=1.15pt
    },
    hz image/.style={
      circle,
      draw=black,
      fill=white,
      line width=0.55pt,
      inner sep=1.3pt
    }
  }

  \providecommand{\Description}[1]{}

  \resizebox{\linewidth}{!}{%
    \begin{tikzpicture}[
      x=1cm,
      y=1cm,
      font=\small,
      line cap=round,
      line join=round
    ]
      \path[use as bounding box]
        (-1.7,-1.25) rectangle (6.65,1.65);

      \begin{scope}
        \node[font=\small\bfseries] at (0.25,1.48)
          {(a) Off target: no solution};

        \draw[hz line] (-1.48,0.65) -- (1.45,0.65)
          node[hz label,right] {$l_1$};
        \draw[hz line] (-1.48,-0.65) -- (1.45,-0.65)
          node[hz label,right] {$l_2$};

        \draw[hz aux,dotted] (-1.4,0.05) -- (1.5,0.05);

        \draw[hz crease] (0.1,-0.9) -- (0.1,0.92);
        \node[hz label,right] at (0.1,-0.28) {$f$};

        \draw[hz aux]
          (0.1,-0.49) -- (0.26,-0.49) -- (0.26,-0.65);

        \draw[hz arrow] (-1,0.05) -- (1.2,0.05);

        \node[hz point] at (-1,0.05) {};
        \node[hz label,below] at (-1,0.05) {$p$};

        \node[hz image] at (1.2,0.05) {};
        \node[hz label,below] at (1.2,0.05) {$r_f(p)$};

        \node at (0.15,-1.12) {$p\notin l_1$};
      \end{scope}

      \begin{scope}[xshift=4.6cm]
        \node[font=\small\bfseries] at (0.18,1.48)
          {(b) On target: not unique};

        \draw[hz line] (-1.48,0.65) -- (1.63,0.65)
          node[hz label,right] {$l_1$};
        \draw[hz line] (-1.48,-0.65) -- (1.63,-0.65)
          node[hz label,right] {$l_2$};

        \draw[hz crease] (-0.35,-0.9) -- (-0.35,0.84);
        \draw[hz crease] (0.2,-0.9) -- (0.2,0.84);

        \node[hz label,left] at (-0.35,-0.18) {$f$};
        \node[hz label,right] at (0.2,-0.18) {$g$};

        \draw[hz aux]
          (-0.35,-0.49) -- (-0.19,-0.49) -- (-0.19,-0.65);
        \draw[hz aux]
          (0.2,-0.49) -- (0.36,-0.49) -- (0.36,-0.65);

        \node[hz point] at (-1,0.65) {};
        \node[hz label,above] at (-1,0.65) {$p$};

        \node[hz image] at (0.3,0.65) {};
        \node[hz label] at (0.3,1.07) {$r_f(p)$};

        \node[hz image] at (1.4,0.65) {};
        \node[hz label] at (1.4,1.07) {$r_g(p)$};

        \node at (0.15,-1.12) {$p\in l_1$};
      \end{scope}
    \end{tikzpicture}%
  }

  \caption{Why Huzita 7 requires nonparallel input lines for unique existence.}

  \label{fig:huzita-seven}
\end{figure}
\paragraph{Huzita 7:} Given one point $p$ and two lines $l_1$ and $l_2$, there is a fold that places $p$ onto $l_1$ and is perpendicular to $l_2$.
\begin{lstlisting}
theorem huzita_7 (p : Point) (l₁ l₂ : Line)
  (h : ¬ parallel l₁ l₂) :
  ∃! f : Fold,
    on_line l₁ (f_places_p f p) ∧ perpendicular f l₂
\end{lstlisting}
Perpendicularity fixes the crease direction, so the displacement from $p$ to its image is parallel to $l_2$, or zero.
For nonparallel input lines, the line through $p$ parallel to $l_2$ meets $l_1$ at a unique point $q$.
If $q\ne p$, their perpendicular bisector is the required crease.
If $q=p$, the perpendicular-bisector construction does not apply; instead the crease must pass through $p$ and be perpendicular to $l_2$, which still determines it uniquely.
The actual coordinate construction handles both cases uniformly, without introducing a hypothesis that $p$ is off its target.
\par
In coordinates, if $l_2$ has normal $(\alpha_2,\beta_2)$, we use $(-\beta_2,\alpha_2)$ as the crease normal.
The placement condition then becomes a linear equation for the constant coefficient.
The coefficient of this unknown is nonzero under nonparallelism.
The construction multiplies by this factor before creating the line, leaving the raw coefficients free of division.
The placement and perpendicularity properties are verified algebraically, and the same equation establishes uniqueness after normalization.
\par
Figure~\ref{fig:huzita-seven} shows why the hypothesis characterizes \emph{unique} existence, rather than existence alone.
For parallel input lines, a point off $l_1$ cannot reach it by reflection across a crease perpendicular to $l_2$.
A point already on $l_1$, however, stays on that line under every such crease, so multiple folds work.
The formal proof exhibits two different creases to refute uniqueness.
The two cases, including coincident input lines, yield \texttt{huzita\_7\_iff}:
\[
    \bigl(\exists! f,\ r_f(p)\in l_1\land f\perp l_2\bigr)\quad
    \Longleftrightarrow\quad l_1\not\parallel l_2.
\]
\section{Objects we made}
In addition to conceiving of the necessary definitions and structures, and creating the Lean framework to describe origami folds and models, we also made several extensions of the basic framework to demonstrate a full pipeline from being able to view models as more traditional fold patterns to being able to prove theorems about given models.
\label{sec:objects}
\subsection{The Crease Pattern Inspector}
\label{ssec:crease_pattern_inspector}
\begin{figure}[]
    \centering
    \includegraphics[width=0.95\linewidth]{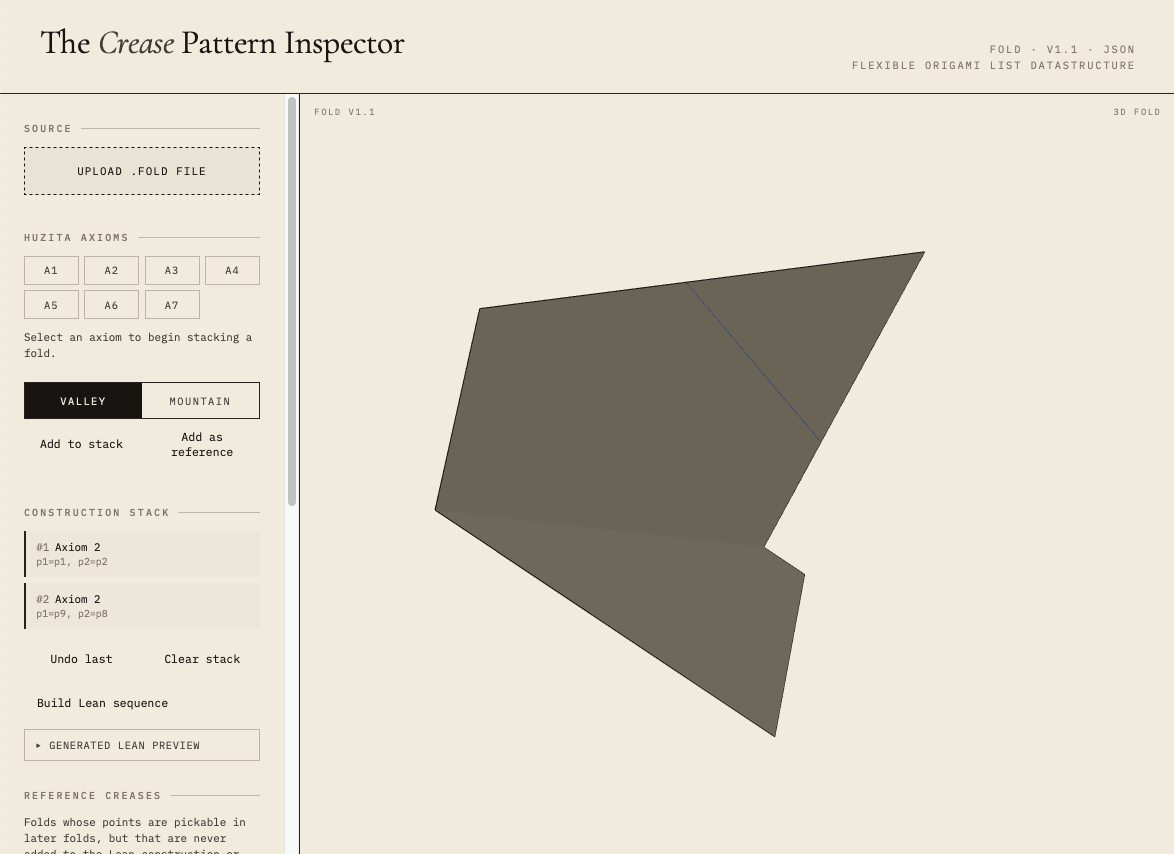}
    \label{fig:crease_inspector}
    \caption{The Crease Pattern Inspector}
\end{figure}
In order to visualize some of the folds that we made, as well as to help in creating new folds and providing newcomers with a more digestible interface with the fold structure, we created a web tool that we have called The Crease Pattern Inspector.

The user is able to construct a 2D crease pattern using the inspector and also view it as a 3D model rendered using a bar-and-hinge model implemented in Rust. We also made the inspector use the same vector based definition of folds mentioned in subsection \ref{ssec:structure}, and treated multiple folds as a stack that can be pushed onto with new folds or popped off of. This means that using a relatively straightforward Python script, the inspector can procedurally convert any model created within it into a representation in Lean, plus a proof of its own viability.
\subsubsection{Imposition of the Huzita Axioms.}
This Crease Pattern Inspector also allows for the validation of the completeness of the Huzita axioms as proved by Robert Lang \cite{Lang1996}. One constraint with Lean (and indeed many other means of formalizing origami) with respect to showing the completeness of the Huzita axioms is that any numerical values can be simply declared, including values not necessarily proven to be constructible with origami. Our solution was to impose restrictions within the inspector interface, wherein we were able to limit all available points and values to those which could be derived or constructed by a combination of the Huzita axioms.

The user can initially only reference the four points that represent the four corners of an unfolded unit square piece of paper and its four corresponding edges. Upon the application of one of the Huzita axioms to create a new fold, this new fold's intersections with the edges of the paper and with other extant folds become "available" to the user, much as the intersection of folds on a piece of paper become defined to the folder.

This was also where we created a notion of a reference crease, wherein a fold can be made that never actually becomes part of the final model, but can make available points that can be used in other folds. This is analogous to say, folding a piece of paper in half, using the corners as reference and then being able to reference the midpoints of two of the sides for other folds even if the fold itself is not part of the final model.

\begin{figure}[]
    \begin{subfigure}[b]{0.3\textwidth}
        \includegraphics[width=\textwidth]{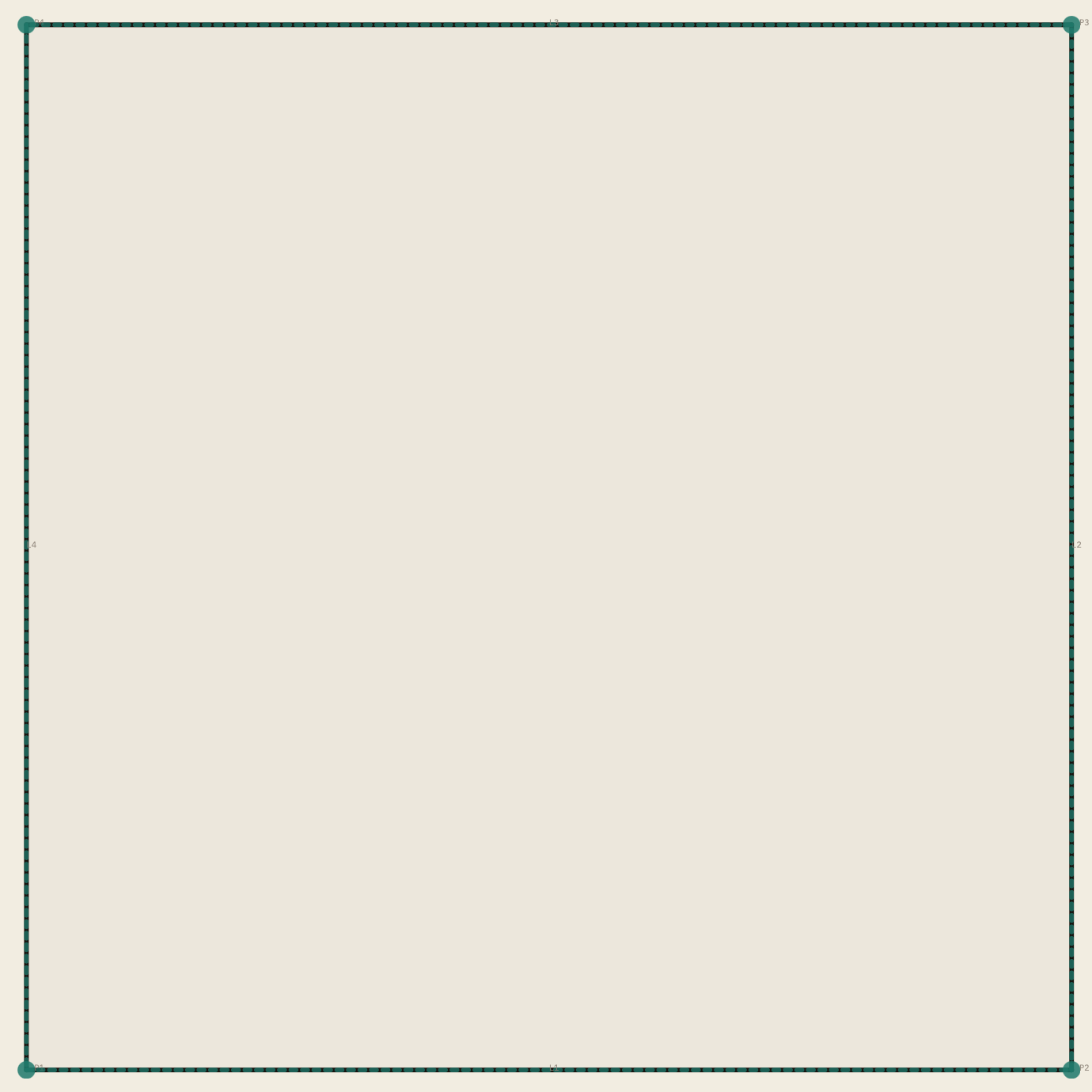}
        \caption{In the initial state, only the four corners of the paper are defined points that are "available".}
    \end{subfigure}
    \par\bigskip
    \begin{subfigure}[b]{0.3\textwidth}
        \includegraphics[width=\textwidth]{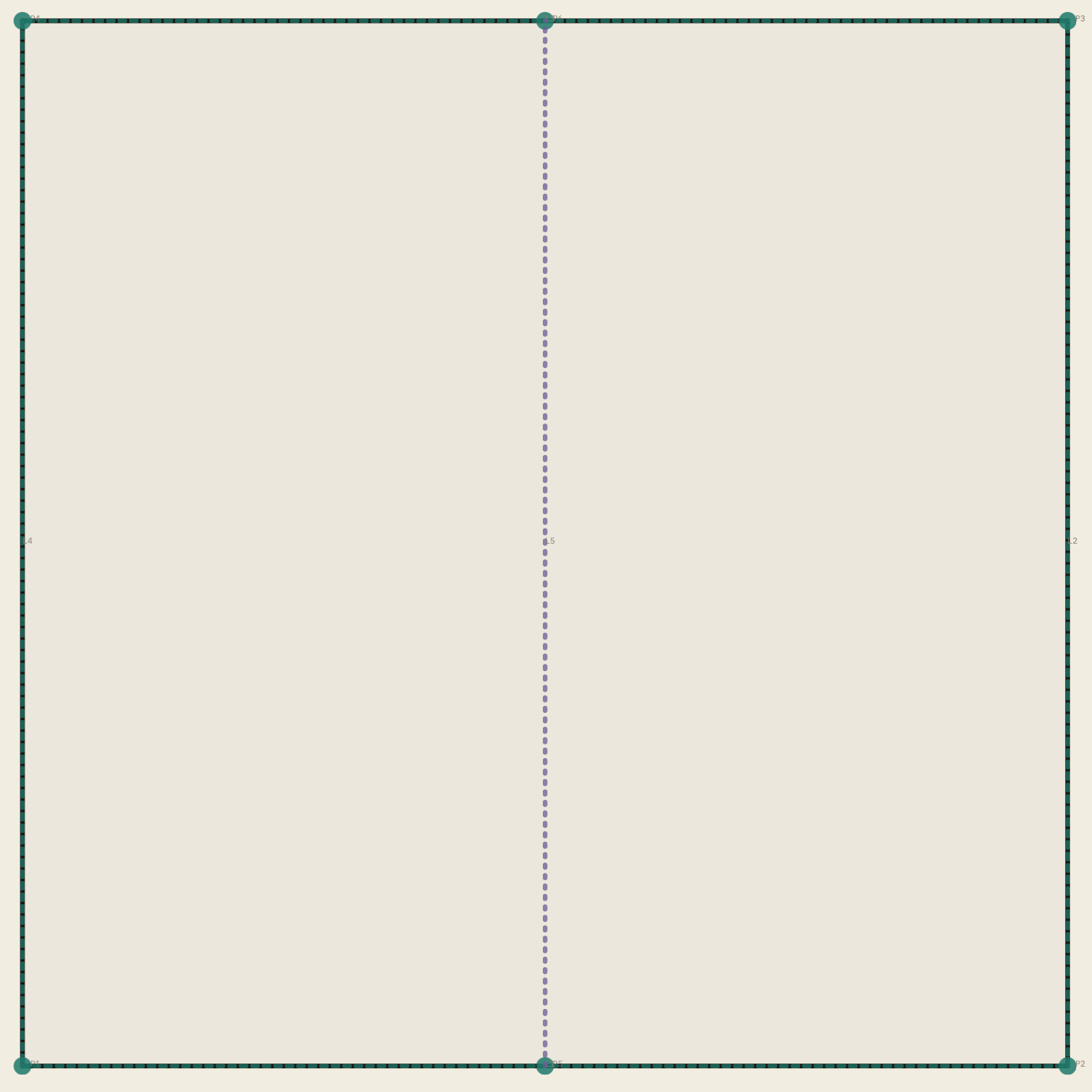}
        \caption{Applying the Second Huzita Axiom to the upper left and right hand corners yields a fold that runs vertically through the center of the paper. This fold's intersections with the edges of the paper become newly defined points.}
    \end{subfigure}
    \par\bigskip
    \begin{subfigure}[b]{0.3\textwidth}
        \includegraphics[width=\textwidth]{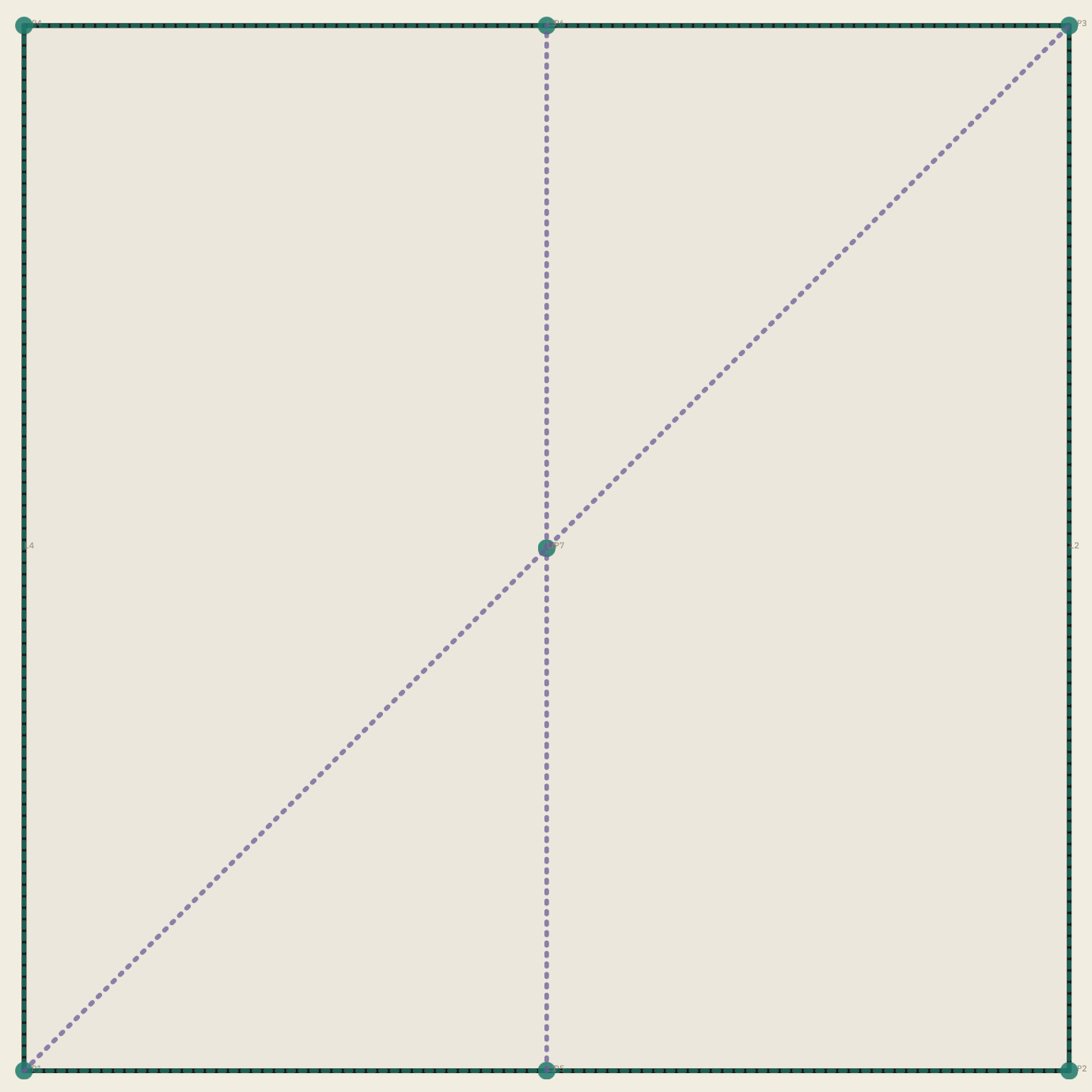}
        \caption{Applying the first Huzita Axiom to the upper right and lower left corner points yields a diagonal fold, whose intersection with the first fold now defines the center point.}
    \end{subfigure}
\label{fig:hagaPoints}
\caption{A brief example of how the Crease Pattern Inspector gates any available points to those that are derived via application of the Huzita Axioms}
\Description[Basic Huzita Construction]{A brief example of ow the Crease Pattern Inspector gates any available points to those that are derived via application of the Huzita Axioms.}
\end{figure}

\subsection{Handcrafted proofs}

Some important geometric problems were proven by Wantzel \cite{wantzel1837} to be unfeasible using compass-and-straightedge. However when provided with origami operations we find ourselves able to achieve such constructions. Two famous examples are the angle trisection and the Delian problem (doubling the cube).

\subsubsection{Angle Trisection}
\label{subsub:tri}
The following construction was discovered by Abe in the 1970s (the interested reader is referred to \cite{Richeson2012}).

\begin{figure}[bthp]
    \centering
    \includegraphics[width=0.95\linewidth]{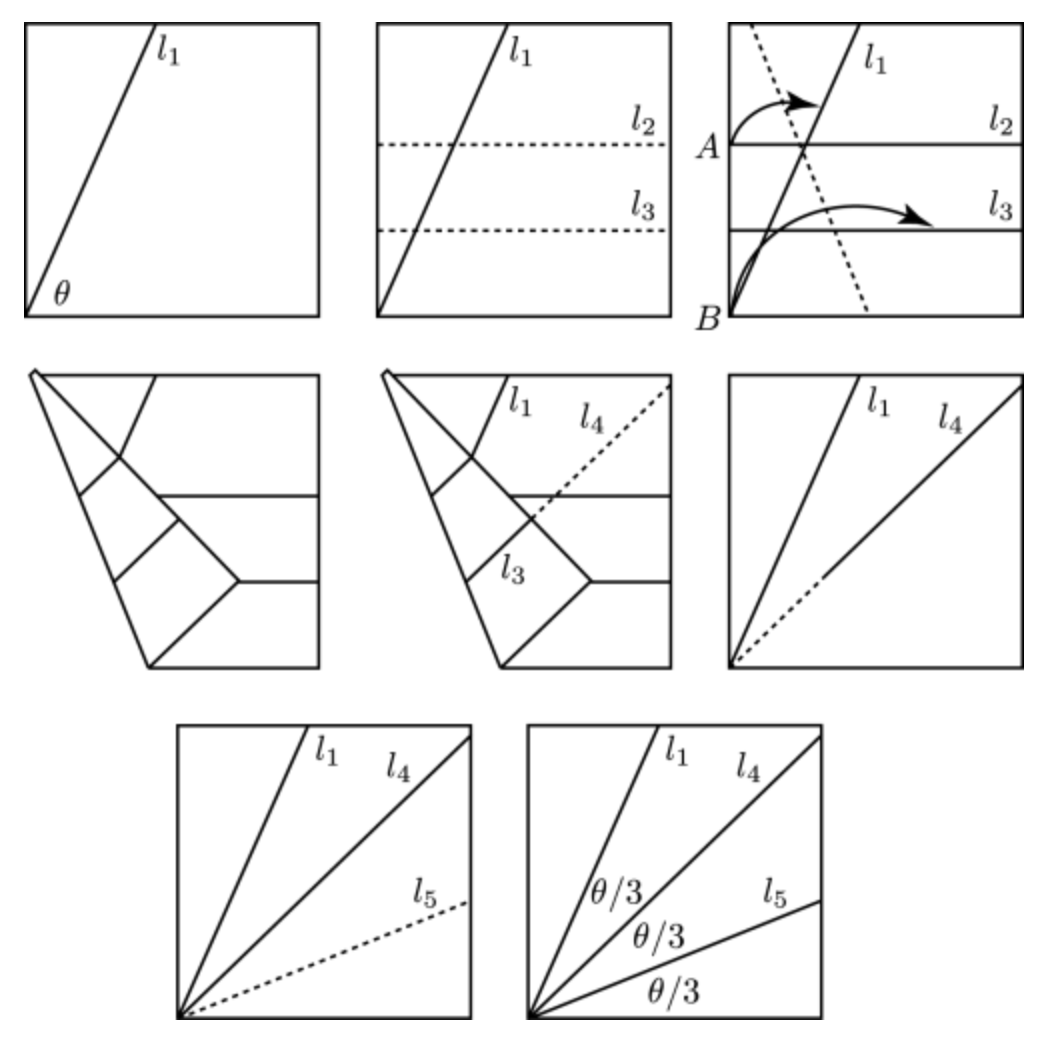}
    \caption{Steps of the trisection using origami \cite{Richeson2012}.}
    \label{fig:trisection_steps}
\end{figure}

\begin{figure}[htbp]
    \centering
    \includegraphics[width=0.95\linewidth]{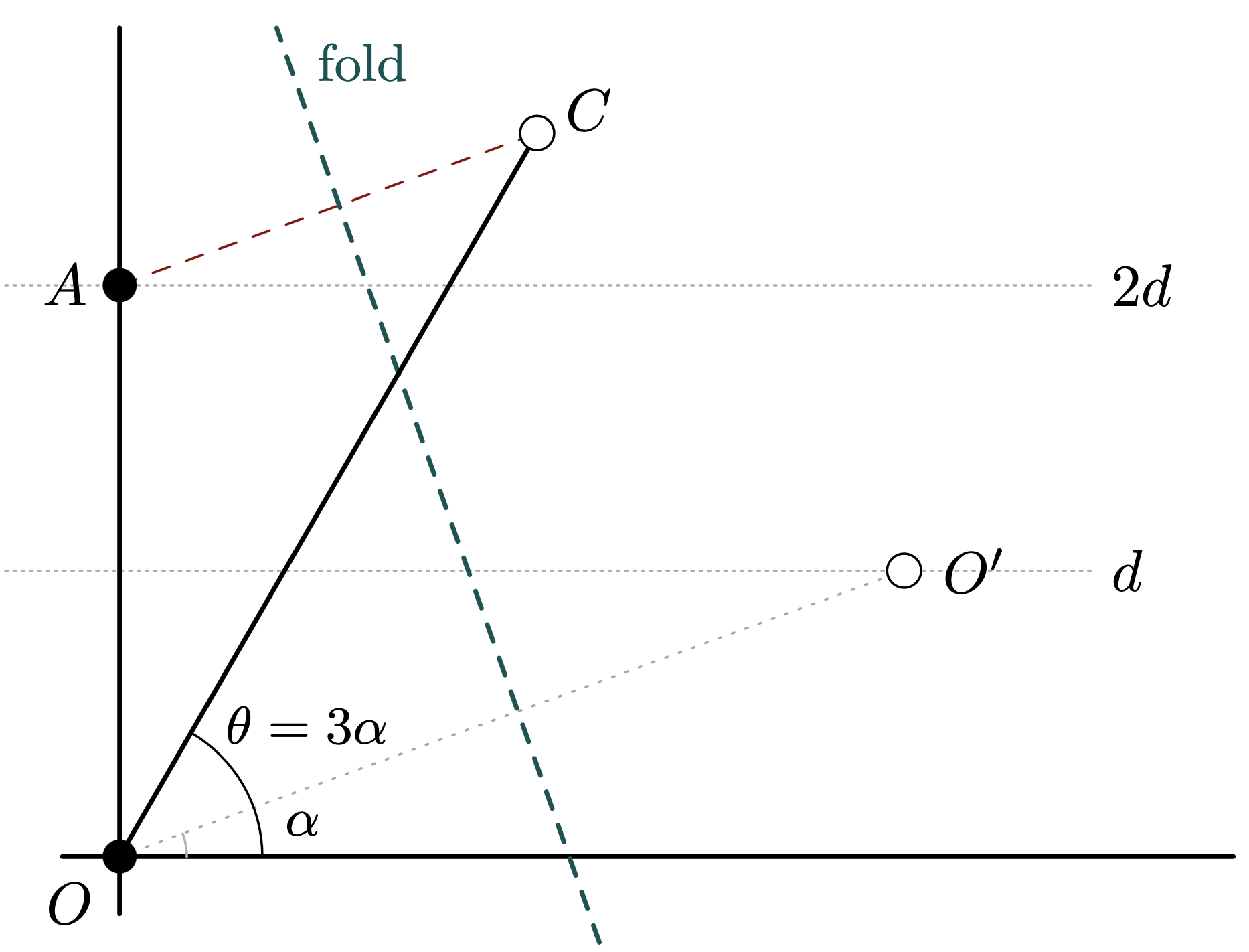}
    \caption{Simplified final state of the trisection process}
    \label{fig:finalTrisection}
\end{figure}

From Fig \ref{fig:finalTrisection} it can be proven that the angle $\theta$ is indeed trisected. This can be done in various ways \cite{BenAri2022Origami, MilojkovicMarinkovic2026}. Different proofs might achieve the same goal and have significantly different formalization complexity depending on the verbosity of the proof or the available API. For instance in this case the reasoning is relatively straightforward when dealing with congruent angles, but is difficult to formalize in Lean due to its geometric nature and the current Mathlib API.

One other way of proving this construction is through numerical considerations, by providing proofs that the construction is valid through different verifications. 

For instance the following proves that C lies on the ray from the origin
at angle $3\alpha$ (by showing that $C_x · sin(3\alpha) = C_y · cos(3\alpha)$).

\begin{lstlisting}
theorem trisection (α d : ℝ) (_hd : 0 < d) (hα : 0 < α) (hα' : α < π / 6) :
    Cx d α * sin (3 * α) = Cy d α * cos (3 * α)
\end{lstlisting}

where we define the point C with coordinates $(C_X, C_Y)$.
\begin{lstlisting}
def Cx (d α : ℝ) : ℝ :=
    d * cos α * (1 - 4 * sin α ^ 2) / sin α
def Cy (d α : ℝ) : ℝ :=
    d * (1 + 2 * cos (2 * α))
\end{lstlisting}

The file AbeTrisection.lean gives proofs of all the required properties (C is indeed what we expect it to be, trisection happens in the expected quadrant, and other properties.)

While this method differs from standard proof methods (we are verifying properties instead of proving a very general theorem), we find it interesting for model testing and fast iteration on given values/ideas, where it might be more valuable to test one \textit{general} example rather than find a generalization.

\subsubsection{Delian Problem}

A somewhat similar approach is used to prove the correctness of a construction for the Delian problem. The difference with the previous example is that we provide here the specific fold involved instead of only using the end points. The proof is also significantly shorter. Consider that we wish to double the volume of the unit cube. This means we need to construct a cube of volume $V = 2 * 1 = 2$, that is a cube with side length equal to $\sqrt[3]{2}$. The problem reduces to being able to construct this number in some way.

\begin{figure}[htbp]
    \centering
    \includegraphics[width=1\linewidth]{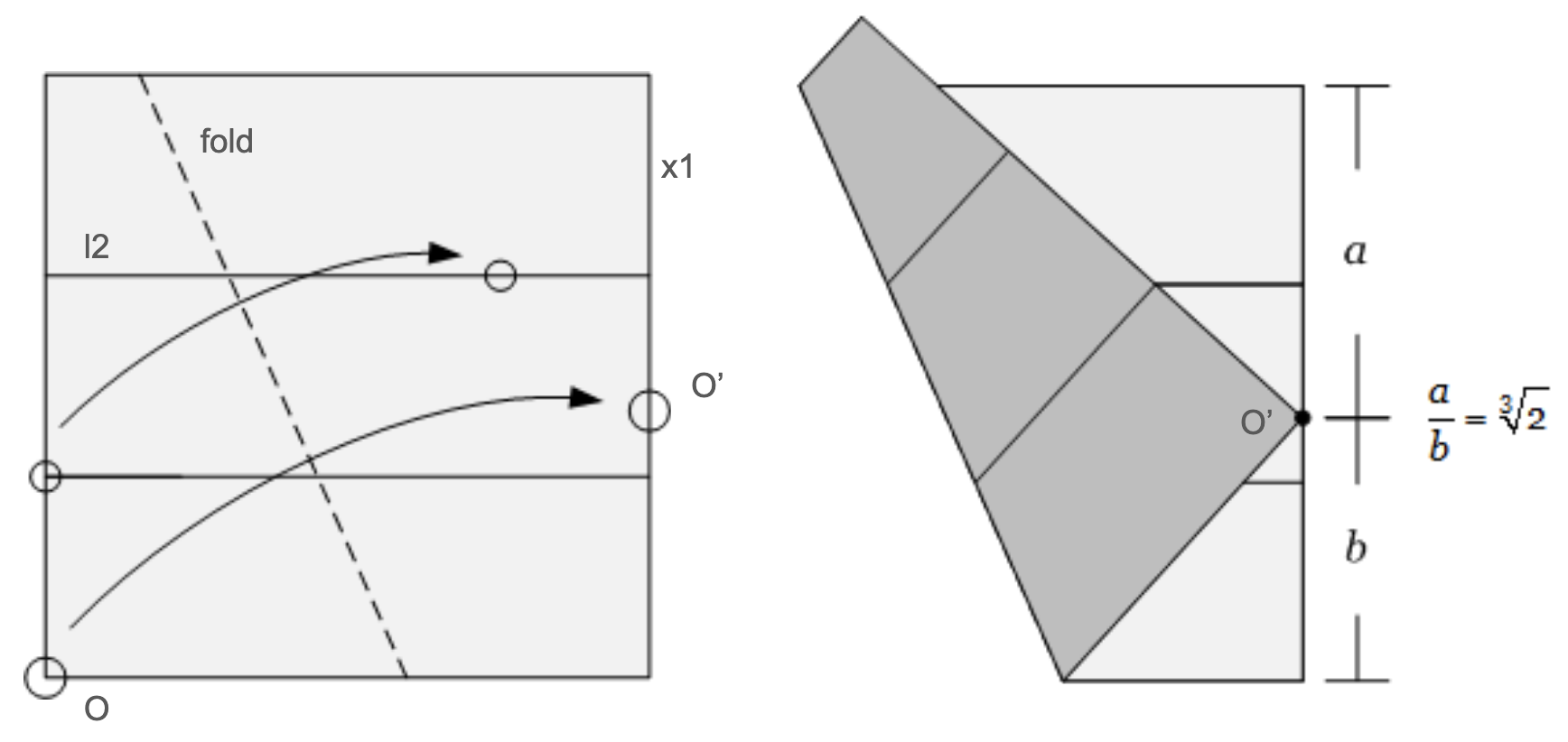}
    \caption{Fold leading to $\sqrt[3]2$}
    \label{fig:delianfold}
\end{figure}

Following the method detailed in \cite{cutoutfoldup_doublecube}, we start by defining l2 the horizontal line at $y = \frac{2}{3}$ and x1 the line at $x = 1$.
Note that creating creases at $y = \frac{1}{3}$ and $y = \frac{2}{3}$ is possible as a result of Haga's theorem (see \ref{ssec:haga}).

\begin{lstlisting}
def l2 : Line := ⟨0, 1, -2/3, by simp⟩
def x1 : Line := ⟨1, 0, -1, by simp⟩
\end{lstlisting}

with the normalized \textit{Line} being defined as in \ref{ssec:structure}.

Then we define A and O
\begin{lstlisting}
def O : Point := ⟨0, 0⟩; def A : Point := ⟨0, 1/3⟩
\end{lstlisting}

and define a fold (the dashed line)

\begin{lstlisting}
def cbrt2 : ℝ := 2 ^ (1/3 : ℝ)
def a : ℝ := 1 + cbrt2
def b : ℝ := 1
def c : ℝ := -(1/6)*((1+cbrt2)^2 + 3)

def f : Fold := ⟨a, b, c, by unfold b; simp⟩
\end{lstlisting}

The values for the fold are manually derived, but so far nothing formally guarantees that the fold is the one we are looking for. Hence we verify that the fold is mapping the points as intended.
\begin{lstlisting}
lemma f_places_O_on_right_border :
  let O' : Point := f_places_p f O
  on_line x1 O'
  
lemma f_places_A_on_l2 :
  let A' : Point := f_places_p f A
  on_line l2 A'
\end{lstlisting}

where for instance f\_places\_p is defined as follows

\begin{lstlisting}
def f_places_p (f : Fold) (p : Point) : Point :=
  let d := (f.a * p.x + f.b * p.y + f.c) / (f.a^2 + f.b^2)
  { x := p.x - 2 * f.a * d
    y := p.y - 2 * f.b * d }
\end{lstlisting}

Since the fold is valid, we can now transform the origin point O through this fold and get the point O'
\begin{lstlisting}
def O' : Point := f_places_p f O
\end{lstlisting}

The rest of the proof simply verifies the desired property.
\begin{lstlisting}
def α : ℝ := 1 - O'.y
def β : ℝ := O'.y

theorem DoublingTheCube : α / β = cbrt2
\end{lstlisting}

\subsubsection{Haga's Theorem}
\label{ssec:haga}
\begin{figure}
    \centering
    \includegraphics[width=0.95\linewidth]{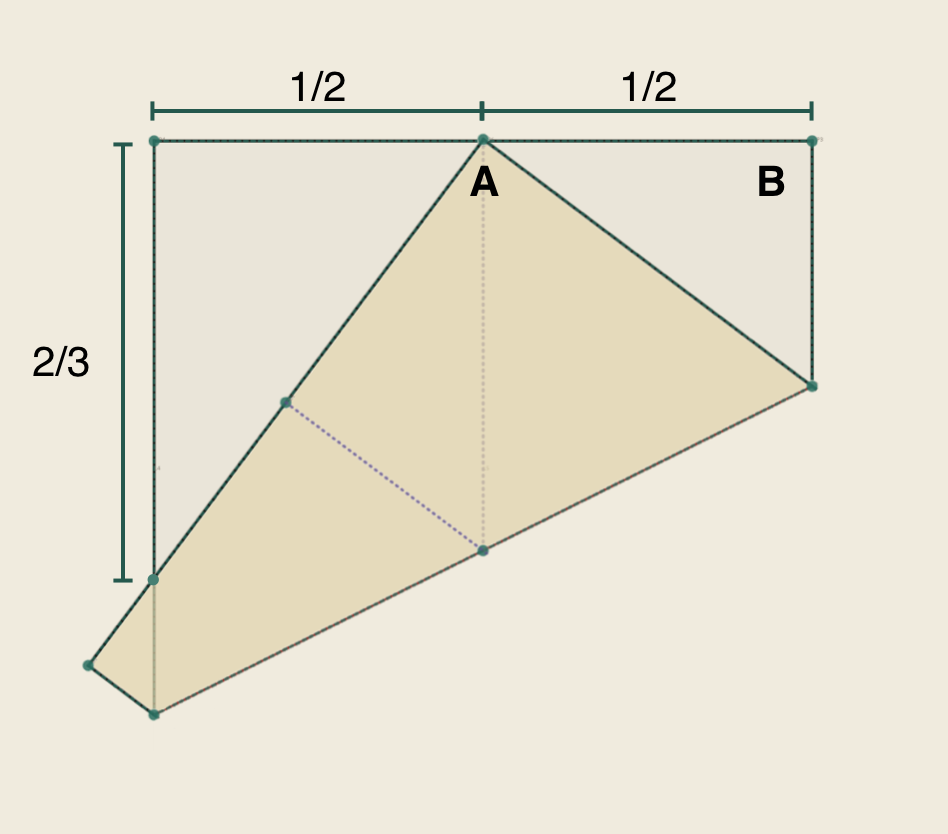}
    \caption{Final state of the $n=2$ case of Haga's Theorem}
    \label{fig:basic_haga_diagram}
\end{figure}

Haga's first theorem \cite{Haga2008}, in its most basic form, dictates that when a corner of a square piece of paper is folded onto the midpoint of the opposite side of the paper, the intersection of the diagonal edge with the unfolded edge of the paper shall be exactly $\frac{1}{3}$ from the corner.

\begin{figure}[H]
    \centering
    \begin{subfigure}[t]{0.2\textwidth}
        \includegraphics[width=\textwidth]{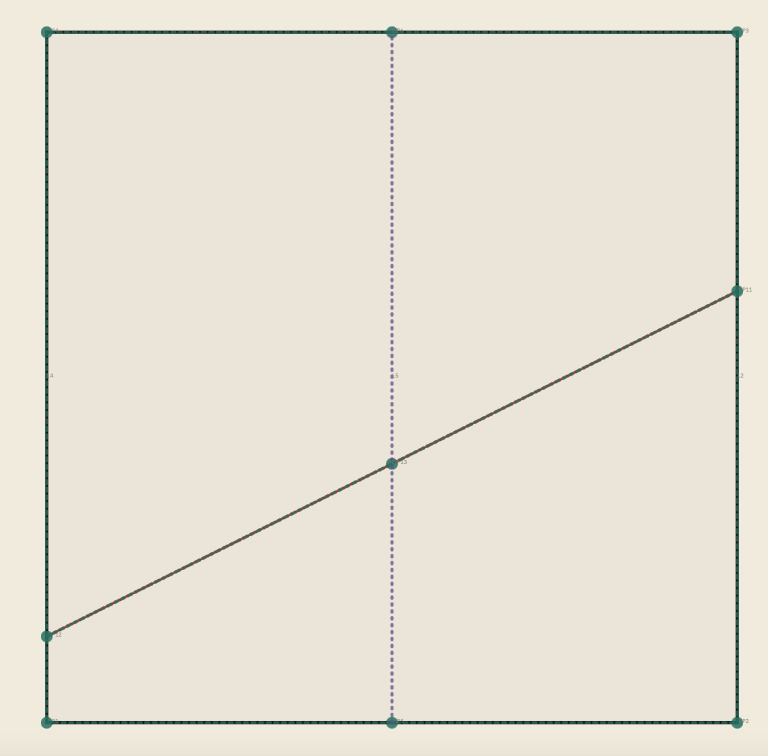}
        \caption{The 2D representation of Haga's theorem in a crease pattern.}
    \end{subfigure}
    \hspace{5mm}
    \begin{subfigure}[t]{0.2\textwidth}
        \includegraphics[width=\textwidth]{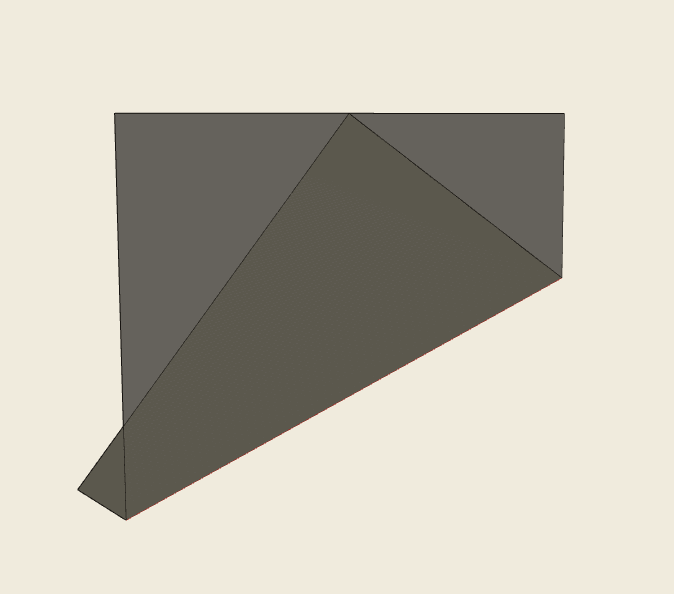}
        \caption{The 3D representation of the Haga's theorem fold.}
    \end{subfigure}
    \caption{The crease pattern corresponding to the n=2 case of Haga's theorem}
    \label{fig:haga_pattern}
\end{figure}

Using the Crease Pattern Inspector, we created the basic crease pattern described above and exported it to Lean, yielding the following structure and theorem. 

\begin{lstlisting}
axiom p1 : Point -- picked at (1, 0)
axiom p2 : Point -- picked at (0.5, 1)
axiom l1 : Line -- crease (0, 0.125) -> (1, 0.625)
axiom p3 : Point -- picked at (1, 0.625)
axiom p4 : Point -- picked at (0, 0.125)
axiom p5 : Point -- picked at (0.5, 0.375)

axiom hyp1 : p1 ≠ p2

theorem haga_construction : True := by
  obtain ⟨f1, h1, -⟩ := huzita_2 p1 p2 hyp1
  trivial
\end{lstlisting}

The above theorem and proof simply state the viability of the Haga fold.
To reason about the actual result of Haga's theorem, we then first created the theorem.

\begin{lstlisting}
theorem haga_first_theorem (crease : Fold) :
  let pA : Point := ⟨1, 0⟩
  let pB : Point := ⟨(1/2 : ℚ), 1⟩
  let _ : Point := ⟨0, 0⟩
  let pLeftIntersect : Point := ⟨0, (1/3 : ℚ)⟩
  is_huzita_2_compliant_fold crease pA pB →
  let lowerEdge : Line := {a := 0, b := 1, c := 0, nontrivial := by simp, normalized := by simp}
  on_line (f_places_l crease lowerEdge) pLeftIntersect
\end{lstlisting}
Which accepts an arbitrary crease, representing our Haga fold, and says that if it maps point $A$ to the midpoint at the top as in Figure ~\ref{fig:basic_haga_diagram}, then the intersection of the bottom of the paper and the left side should be $\frac{1}{3}$ from the bottom.

We then utilized a proof from the British Origami Society \cite{WalkerBOM2025} to calculate the actual line coefficients, giving us a fold that we can define explicitly in terms of $x-2y+\frac{1}{4}=0$, which we call \lstinline{alsoCrease}. The second Huzita axiom has an existence assertion that proves the viability of the Haga fold in \lstinline{haga_construction} above, but does not necessarily give us the exact line. And this time we used the uniqueness property of the axiom to assert that the constructed \lstinline{alsoCrease} fold is equivalent to the original Haga fold \lstinline{crease} because it also maps point $A$ as \lstinline{crease} does.
\begin{lstlisting}
    let alsoCrease : Fold := {a := 1, b := -2, c := 1/4, nontrivial:=by simp, normalized:=by simp}
    let alsopB : Point := f_places_p alsoCrease pA
    have pBEquiv : (alsopB = pB) := by
      unfold alsopB f_places_p alsoCrease pA pB
      simp [is_huzita_2_compliant_fold, pA, pB] at *
      grind
    have creaseEquiv : alsoCrease = crease := by       
        have h_combined : f_places_p alsoCrease pA = f_places_p crease pA := by
            rw [←pBEquiv] at h
            grind[is_huzita_2_compliant_fold]
          have hne : pA ≠ pB := by
            intro hEq
            have hy := congrArg Point.y hEq
            unfold pA pB at hy
            norm_num at hy
          apply huzita_2_uniqueness alsoCrease crease pA pB hne
          trivial
\end{lstlisting}

Because we now know the exact line defined by the fold, it becomes a simple matter of computation to determine the line defined by the edge folded over the fold, and therefore where it intersects with the opposite edge. We can then use Lean's built in \lstinline{simp} and \lstinline{norm_num} tactics for simplifying basic algebra and arithmetic.

To tie this back to the Haga fold pattern generated by the Crease Pattern Inspector, we restated the original existence theorem, but with \lstinline{is_huzita_2_compliant_fold f p1 p2} included for later use
\begin{lstlisting}
theorem haga_fold_exists : ∃ f : Fold, is_huzita_2_compliant_fold f p1 p2 := by
  obtain ⟨f1, h1, -⟩ := huzita_2 p1 p2 hyp1
  exact ⟨f1, h1⟩
\end{lstlisting}
from there, we are able to feed this into \lstinline{haga_first_theorem} to close the proof using the generated model.
\begin{lstlisting}
theorem haga_result :
  let pLeftIntersect : Point := ⟨0, (1/3 : ℚ)⟩
  let lowerEdge : Line := {a := 0, b := 1, c := 0, nontrivial := by simp, normalized := by simp}
  on_line (f_places_l (Classical.choose haga_fold_exists) lowerEdge) pLeftIntersect
\end{lstlisting}
This was all to prove the $n=2$ case of Haga's theorem, but it can also be generalized to a general $\frac{1}{n}$ case. We also create a standalone proof that uses a similar definition to the base case proof to prove the induction step that if for $n\in \mathbb{N}, n\geq 2$, $\frac{1}{n}$ is constructible using paper folds, then $\frac{1}{2n-1}$ is therefore also constructible (note that in the $n=2$ case, that since $\frac{1}{2}$ is constructible by folding the paper in half, the original proof proves the ability to construct $\frac{1}{2*2-1}=\frac{1}{3}$). This can be used in the future for an induction proof saying that $\forall n\in \mathbb{n}$, that through a sequence of folds, $\frac{1}{n}$ can be constructed with the full induction step:
\begin{itemize}
    \item Case 1: $n+1$ is even, we can then use the fact that folding in half is trivially possible to get and that $\frac{2}{n+1}$ is constructible in our induction hypothesis to imply $\frac{1}{n+1}$ is constructible
    \item Case 2: $n+1$ is odd, which means $\exists n'\in \mathbb{N}, n+1=2n'-1$ meaning that $\frac{1}{n'}$ being constructible $\implies \frac{1}{n+1}$ is constructible
\end{itemize}

This also yields the ability to construct arbitrary rational numbers, which is another way to approach \ref{ssec:origamiConstructibleNums} and therefore $\mathbb{Q}\subseteq \mathbb{O}$.

\subsection{The Origami-Constructible Numbers}
\label{ssec:origamiConstructibleNums}

Very interesting results in origami theory stem from viewing the piece of paper as the complex plane \cite{king2004origami, Alperin2000}.
We formalize its structure in Lean.

\begin{lstlisting}
inductive Origami : ℂ → Prop
  | one : Origami 1
  | add {z w} :
        Origami z → Origami w → Origami (z + w)
  | neg {z} : Origami z → Origami (-z)
  | mul {z w} :
        Origami z → Origami w → Origami (z * w)
  | inv {z} : Origami z → Origami z⁻¹
  | sqrt {z w} : Origami z → w ^ 2 = z → Origami w
  | cbrt {z w} : Origami z → w ^ 3 = z → Origami w

\end{lstlisting}

Note that none of these properties is \textit{obvious} or axiomatic, especially the last three points. However, we decided to consider them as properties of $\mathbb{O}$ for the scope of this section. Future work will focus on proving these points from geometric constructions as in \cite{king2004origami}.

\begin{lstlisting}
def OO : Subfield ℂ where
  carrier := {z | IsOrigami z}
  ...
  
theorem ℚ_mem_OO : ∀ q : ℚ, (q : ℂ) ∈ OO := by simp
\end{lstlisting}

Since doubling the cube is reducible to constructing $\sqrt[3]{2}$, we provide a proof that $\sqrt[3]{2} \in \mathbb{O}$. From our previous definition, we could give the following proof

\begin{lstlisting}
theorem cbrt_two_mem {z : ℂ} (hz : z ^ 3 = 2) : z ∈ OO := by 
  have h2 : 2 ∈ OO := by simp
  exact IsOrigami.cbrt h2 hz
\end{lstlisting}

This feels like a tautology since we take
\begin{lstlisting}
Origami z → w ^ 3 = z → Origami w 
\end{lstlisting}
for granted. We choose to prove this in another way which requires proving Cardano's formula, stating that every monic cubic with origami coefficients has a root in $\mathbb{O}$.
\begin{lstlisting}
theorem exists_cubic_root_mem {a b c : ℂ}
    (ha : a ∈ OO) (hb : b ∈ OO) (hc : c ∈ OO) :
        ∃ s ∈ OO, s ^ 3 + a * s ^ 2 + b * s + c = 0 
\end{lstlisting}

This result proves that origami construction can go further than the limitation of compass-and-straightedge where every problem that leads to an irreducible equation whose degree is not a power of 2, cannot be solved \cite{wantzel1837}.
That $\sqrt[3]2$ can be constructed is now proven in a more elegant way using exists\_cubic\_root\_mem.
\begin{lstlisting}
theorem cbrt_two_mem {z : ℂ}
    (hz : z ^ 3 = 2) : z ∈ OO
\end{lstlisting}

\section{Conclusion and Future Work}
\label{sec:conclusion}
In undertaking this project, we have created a novel framework that allows for a mathematical representation of origami folds in Lean, as well as a pipeline from a user-friendly visual crease pattern tool in our Crease Pattern Inspector. We have also extended and utilized this framework by formalizing several key existing results in origami folding, including the Huzita operations, Angle Trisection, Delian's Problem, the origami-constructible numbers $\mathbb{O}$, and Haga's theorem. In doing so, we have demonstrated the viability and usefulness of the Lean language as a way to represent, validate, and reason about the characteristics of different paper folds.

In the future, we hope to explore the topology of more complex paper folds, including how multiple intersecting folds interact with each other and how different orderings of a given set of folds affect the characteristics and viability of the model. From this, we hope to proceed to exploring non-flat folding and modular origami wherein multiple models can be joined together.

\begin{acks}
We thank Henry Yuen, Kunal Marwaha, and Natalie Parham for interesting conversations and constructive feedback.
\end{acks}

\section*{Ethics and Privacy Statement}
\label{sec:ethics}
Pursuant to the ACM Code of Ethics, we have taken steps to ensure that the work presented here represents a good faith effort to advance the field of formal verification of proofs and programs, and we have made sure to be transparent about our methods and the work of others that we may have gained inspiration from. The nature of our work for this project did not necessitate the collection of data, private or otherwise, from any research participants, and thus we do not have any substantive privacy concerns.

\section*{AI Statement}
\label{sec:ai}
We used Claude Code to assist with formalizing and proving the angle trisection and Cardano’s formula. It materially affected parts \ref{subsub:tri} and \ref{ssec:origamiConstructibleNums}. We also used Gemini to assist in finding certain tactics and methods for the primary theorems for both the base and general cases of Haga's theorem, although the basic structure of those proofs was created by hand. Additionally, we used Claude Code to help bridge the gap between the Haga fold generated by the inspector and the proof of Haga's theorem. This would have covered the material in subsection \ref{ssec:haga}.  

We also made extensive use of Claude Code for the server backend and front-end interface of the Crease Pattern Inspector, including all the work and material we covered in subsection \ref{ssec:crease_pattern_inspector}. However, the conception of the inspector, its basic structure, and final validation was performed by hand. Additionally, all the models we created using the inspector were also created by hand.

AI was used for document anonymization and verified by the author. The authors hand-verified the correctness and originality of all content including references.

\bibliographystyle{ACM-Reference-Format}
\bibliography{sample-base}

\end{document}